\documentclass[fleqn,usenatbib,useAMS]{mnras}

\usepackage{booktabs, multirow} % for borders and merged ranges
\usepackage{soul}% for underlines
\usepackage{xcolor,colortbl} % for cell colors
\usepackage{changepage,threeparttable} % for wide tables
\usepackage{graphicx}	% Including figure files
\usepackage{amsmath}	% Advanced maths commands
\usepackage{amssymb}	% Extra maths symbols
\usepackage{multicol}        % Multi-column entries in tables
\usepackage{bm}		% Bold maths symbols, including upright Greek
\usepackage{pdflscape}	% Landscape pages
\usepackage{comment}
\usepackage{appendix}

\usepackage[T1]{fontenc}
\usepackage{ae,aecompl}

\usepackage{newtxtext,newtxmath}
\title[MNRAS \LaTeX\ guide for authors]{
Angular momentum drives proton-rich nucleosynthesis in hyperaccreting neutron stars in common envelopes}

\author[A. D. Hall-Smith]{
Alexander D. Hall-Smith$^{1,3}$
\thanks{Contact e-mail: \href{mailto:alexander.hall-smith@york.ac.uk}{alexander.hall-smith@york.ac.uk}},
S. E. D. Abrahams$^{1,3}$,
A. M. Laird$^{1,3}$,
C. Aa. Diget$^{1}$,
C. Fryer$^{2,3}$,
S. W. Jones$^{2,3}$
\\
$^{1}$School of Physics, Engineering and Technology, University of York, Heslington, York YO10 5DD, UK
\\
$^{2}$ Center for Theoretical Astrophysics, Los Alamos National Laboratory, Los Alamos, NM 87545, USA
\\ 
$^{3}$ NuGrid Collaboration\thanks{ http://nugridstars.org}}

\date{Last updated 2020 June 10; in original form 2013 September 5}

\pubyear{2026}
\usepackage{float}
\begin{document}
\label{firstpage}
\pagerange{\pageref{firstpage}--\pageref{lastpage}}
\maketitle

% Abstract of the paper
\begin{abstract}
\label{abstract}

Interacting binaries can produce a wide range of exotic systems, including X-ray binaries and merging neutron stars, through a mass transfer phase called Common Envelope (CE) evolution. A CE phase can occur during rapid expansion as a star as it moves off the main sequence. If the engulfed star is a compact object (e.g. neutron star), a CE phase can lead to hyperaccretion onto the neutron star. Previous work focused on systems in which the accreting material has low angular momentum, studying turbulent outflows. This study investigates the impact of angular momentum on accreting material leading to the formation of an accretion disk. Disk accretion systems lead to very different nuclear burning conditions. 

This paper presents the results of nucleosynthesis modelling of material ejected from an accretion disk surrounding a 1.5 M$_{\odot}$ neutron star in a CE with a 15 M$_{\odot}$ companion. As material is accreted towards the neutron star, sufficient heating will occur to eject a fracton of the material back into the surrounding envelope, producing a nucleosynthetic yield signature that differs from other explosions. We find that significant mass fractions of rp-process products are synthesised, thereby providing another mechanism for rp-process contribution to galactic chemical evolution, following ejection of the CE. Furthermore, later stages of the CE evolution the accrete helium leading to alpha-rich, supernova-like nucleosynthesis, producing $^{44}$Ti and $^{56}$Ni. Further work on modelling both the accretion disk wind, and the companion envelope ejection, is vital to understand the contributions of these scenarios to chemical evolution.

\end{abstract}

% Select between one and six entries from the list of approved keywords.
% Don't make up new ones.

%%%%%%%%%%%%%%%%%%%%%%%%%%%%%%%%%%%%%%%%%%%%%%%%%%

%%%%%%%%%%%%%%%%% BODY OF PAPER %%%%%%%%%%%%%%%%%%

% The MNRAS class isn't designed to include a table of contents, but for this document one is useful.
% I therefore have to do some kludging to make it work without masses of blank space.
\begingroup
\let\clearpage\relax
\tableofcontents
\endgroup
\newpage

\section{Introduction}
\label{sec:Intro}

Common envelope (CE) evolution occurs when one of the stars in a binary systems begins to expand (typically after hydrogen- or helium-burning exhaustion).  During this expansion, the star can drive such rapid mass transfer on its companion that the companion is overwhelmed, and the system ultimately develops an atmosphere that surrounds (or envelops) both the core of the expanding star and its companion~\citep[for a review, see][]{2013A&ARv..21...59I,Ivanova_CE_evolution}. The frictional drag caused by this envelope leads to the loss of angular momentum in the core-companion orbit, tightening the binary. Many of the close binaries observed in the Universe will go or have gone through a common envelope phase and this phase is critical in the formation of everything from X-ray binaries and recycled pulsars~\citep{2008MNRAS.386..553I} to the progenitors of type I supernovae~\citep{2017MNRAS.469.4763M} and gamma-ray bursts~\citep{Fryer_1999}.  

For massive star binaries, systems can go through more than one common envelope phase.  First, when the more-massive star evolves off the main sequence.  This tightens the binary, making the remnant more likely to remain bound after the supernova explosion. When the companion then evolves off the main sequence, it can envelope the compact remnant companion, causing it to inspiral through the envelope. The accretion onto this compact object as it inspirals through the companion can drive outflows.  If the neutron star spirals into the core, the angular momentum of the core is likely to produce a rapidly accreting disk that may produce a gamma-ray burst~\citep{1998ApJ...502L...9F,2001ApJ...550..357Z,Fryer_2013}, a subclass of gamma-ray bursts known as ultra-long duration gamma-ray bursts~\citep{2011Natur.480...72T,Neights_2025_GRB} or fast blue optical transients~\citep{2025arXiv251009745K}.  Even if the compact object does not spiral into the stellar core, it is likely that these systems form jets that drive outflows with supernova-like outbursts~\citep{2022MNRAS.515.5473R}.  A growing number of multi-dimensional simulations have studied the accretion and its outflow~\citep[e.g.][]{2022MNRAS.514.3212H,2024A&A...691A.244V,2025OJAp....8E..60S,2025PASA...42...76J}.

Knowing that these mergers eject mass that has been processed through close interactions with a neutron star, it is important to understand the detailed yields from these mergers.  Based on the properties of these accretion disks~\citep{Popham_1999}, \citet{2006ApJ...643.1057S} calculated the yields, finding that only in the highest-accretion rates could these disk produce considerable r-process elements.  Several nucleosynthetic studies have studied non-rotating~\citep{Fryer_2006,Keegans_2019,anninos2025rprocessnucleosynthesishyperaccretingneutron} accretion scenarios.  In a non-rotating system, the material accreting onto the neutron star is expected to develop an unstable entropy profile, driving vigorous convection~\citep{1989ApJ...346..847C,1993ApJ...411L..33C} that could produce outflows~\citep{Fryer_2006}.  The temperatures achieved near the surface of the neutron are sufficiently high to drive strong nuclear burning.  \citet{Keegans_2019} studied the yields of these outflows, assuming that the time spent at extreme temperatures and densities was limited to a single turnover time.  The minimal time spent at these extreme conditions limits the type of yields that can be produced.  In detailed hydrodynamic models, \citet{anninos2025rprocessnucleosynthesishyperaccretingneutron} found that, in the complex convective motion near the neutron star, the ejected material can spend much more time in extreme, nuclear-burning conditions.

At extremely high accretion rates, neutrino emission and absorption can be important. Except for the simulations of \citet{Fryer_2006}, these calculations neglected the role of neutrino absorption in setting the electron fraction. In supernovae, this neutrino absorption tends to reset the electron fraction~\citep{2005NuPhA.758...27F,2023ApJ...957L..25S} leading to ejecta with electron fractions near 0.5. As the material expands out of Nuclear Statistical Equilibrium, there is little time for burning pathways with long wait times and most of the ejecta is limited to iron peak elements with only trace amounts of elements produced with atomic weights heavier than 90-100.

Angular momentum can dramatically change this picture. Given the expected accretion of angular momentum, it is likely that disks form, changing the temperature and density evolution of the accreting material.  Material accreting through a disk can spend much more time at sufficiently high temperatures for nuclear burning, leading to the production of heavy elements that can not be produced in a single convective turnover time. Some of this material will be ejected out of the accretion disk. Many disk studies have focused, like \citet{2006ApJ...643.1057S}, on the r-process yield from neutron-rich ejecta.  This requires the material to be held at extreme densities/temperatures for sufficient time for electron capture to deleptonize the material.  As we shall see here, and as discussed in preliminary results by \citet{2023EPJWC.27910002A}, synthesis of proton-rich material is more common in many common envelope cases.  

The scenario investigated in this study consists of a 15 M$_{\odot}$ massive star, at the end of core hydrogen burning, as it expands to engulf a 1.5 M$_{\odot}$ neutron star and forms a common envelope system. As the neutron star inspirals towards the  core of its massive-star companion, it will accrete material and move from the hydrogen-rich outer shell towards the helium-rich core. This allows for a wider range of nucleosynthesis processes than originally considered by \citet{Keegans_2019} - which modelled the neutron star only accreting solar abundance material. 

\section{Neutron star accretion disk trajectories}
\label{sec:Trajectory_dev}

Depending upon the amount of angular momentum in accreting material, the nature of the accretion and mass ejection of a neutron star in a common envelope can behave very differently.  Yield calculations require ``trajectories'' that describe the evolution of the density and temperature with time. If the angular momentum is negligible, convection produces single~\citep{Keegans_2019} or multiple~\citep{anninos2025rprocessnucleosynthesishyperaccretingneutron} cycles to describe the trajectory evolution.  Density and velocity gradients in massive stars will lead to a modest amount of angular momentum accreted in the flow.  In this paper, we focus on models where this angular momentum is sufficient to form a disk (e.g. $5\times10^{16-17}~{\rm \, cm^2~ s^{-1}}$).  The trajectory evolution for this disk scenario can be described in 3 phases:  infall onto the disk, disk evolution, and disk-wind ejection.

For the infall phase, we assume the material accretes at free-fall.  In this scenario, the infall velocity is set to the free-fall velocity ($v_{\rm ff}$):
\begin{equation}
    v_{\rm ff} = (2 G M_{\rm NS}/r)^{1/2}
\end{equation}
where $r$ is the infall radius, $G$ is the gravitational constant and $M_{\rm NS}$ is the neutron star mass.  We assume adiabatic contraction and the material heats up as it falls inward.  This phase persists until the angular momentum in the material halts the infall, causing it to hang up in a disk.  The material hangs up at a radius ($r_{\rm disk}$) when the centrifugal force equals the gravitational force:
\begin{equation}
    r_{\rm disk} = j_{\rm ang}^2/(GM_{\rm NS})
\end{equation}
where $j_{\rm ang}$ is the specific angular momentum of the accreting material.  We assume angular momentum conservation in the collapse so $j_{\rm ang} = 10^{17} {\rm \,cm^2~s^{-1}}$ corresponds to a $10\,{\rm km~ s^{-1}}$ asymmetry at $10^{11}{\rm \, cm}$.

Disk evolution is approximated by an $\alpha$ disk model~\citep{ShakuraSunyaevdisk}.  We assume the accretion rate in the disk is set by the infall accretion rate (e.g. Bondi-Hoyle-Lyttleton or a reduced Bondi-Hoyle accretion~\citep{Bondi52,1995A&A...295..108R,everson20}).  With the $\alpha$ disk prescription, we can estimate the accretion timescale:
\begin{equation}
    t_{\rm disk} = P_{\rm disk}/\alpha
\end{equation}
where $P_{\rm disk} = 2 \pi r_{\rm disk}^{3/2}(G M_{\rm NS})^{-1/2}$ and $\alpha=0.01-0.1$.  The $\alpha$ is a simplified parameterization of the viscous forces in the disk that transport angular momentum and eject matter from the disk.  For material accreting onto to a 1.4\,M$_\odot$ neutron star with an angular momentum, $j_{\rm ang} = 10^{17} {\rm~cm^2~s^{-1}}$, the disk initially hangs up at $r_{\rm disk} = 50\,{\rm km}$ with a disk accretion timescale of 1.8-18s (it can be 125 times longer if the angular momentum is 5 times larger).  As this time is much shorter than the evolution in the common envelope, we can assume it is instantaneous.   

Although the accretion time is short when compared to the evolution of a common envelope, it is long compared to most convective-burning timescales. The trajectories explored in \citet{Keegans_2019} and \citet{anninos2025rprocessnucleosynthesishyperaccretingneutron} operate on the timescale of milliseconds, while trajectories here last over 2000s as a result of inclusion of angular momentum. The timescale on which accretion, disk evolution and ejection occur are also an order of magnitude longer than typical X-ray burst (XRB) timescales. While XRB can reach peak temperatures of 1-1.9GK~\citep{2008ApJS..178..110P}, models explored here can reach peak temperatures of 8GK. For the structure of the disk, we leverage the analytic models by \citet{Popham_1999}.  

Despite many disk calculations, the exact nature of disk accretion and subsequent disk wind remains a matter of active study.  Some models argue that the viscosity can drive mass loss across the entire disk, others argue that the mass ejection occurs primarily at the inner disk radius~\citep{kaltenborn23}.  For our models, we typically assume the wind ejecta is uniformly distributed across the disk.  When the material is blown off in a wind, it accelerates roughly to the escape velocity of the material.  For this final phase of our trajectory, we assume an exponential + power-law evolution.  For explosive supernova and r-process yields, typical trajectories invoke either acceleration or free-streaming regimes.  As the ejecta accelerates, the material slowly expands, causing the density ($\rho$) to drop, following an exponential:
\begin{equation}
    \rho = \rho_0 e^{-t/\tau}
    \label{eq:rhoexp}
\end{equation}
where $\rho_0$ is the density at time 0, and $\tau$ is the dynamical timescale related to the acceleration.  If we further assume the entropy ($S$) is constant and radiation pressure dominates, $S \propto T^3/\rho$, the temperature ($T$) is given by:
\begin{equation}
    T = T_0 e^{-t/(3 \tau)}
    \label{eq:texp}
\end{equation}
where $T_0$ is the temperature at time 0.  Once the velocity accelerates to its peak velocity, we assume the ejecta moves outward at a constant velocity (producing the homologous outflows used in supernova light-curves).  With this assumption, the density and temperature follow power law trajectories:
\begin{equation}
    \rho=\rho_0 (t/\tau)^{-3},
\end{equation}
\begin{equation}
    T=T_0 (t/\tau)^{-1}
\end{equation}
where $\tau$ is the expansion timescale based on the ejecta velocity.  We transition from an exponential to power-law expansion at 3 times this expansion timescale.

\begin{figure}
    \centering
    \includegraphics[width = \columnwidth]{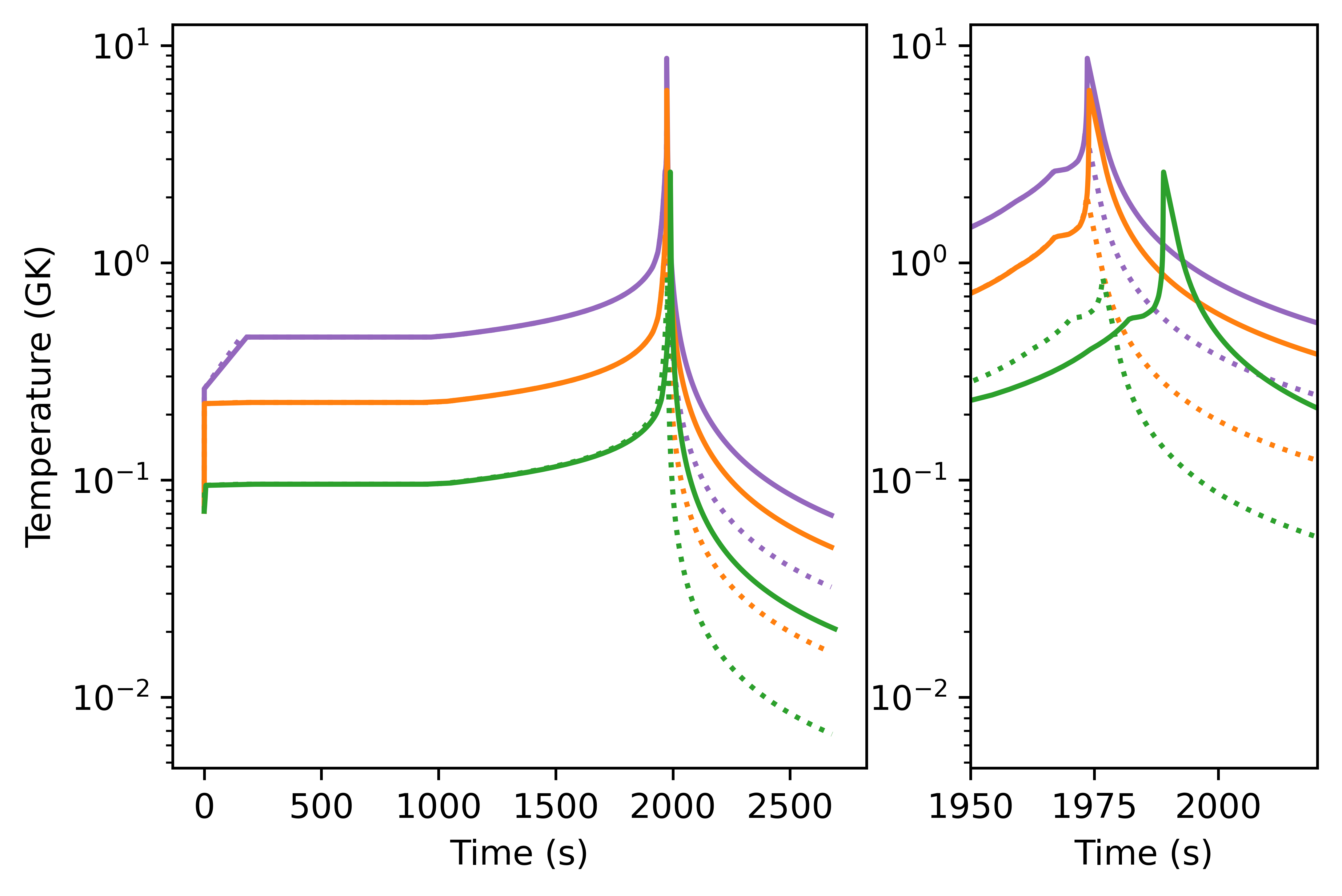}
    \caption{The temperature evolution of three different accretion rate trajectories; $1\times10^{-5} $ M$_{\odot}~$s$^{-1}$ (green), $32\times10^{-5} $ M$_{\odot}~$s$^{-1}$ (orange) and $512\times10^{-5} $ M$_{\odot}~$s$^{-1}$ (purple). These are split into three phases; the infall phase (rapid increase at $t$ = 0), the disk evolution (plateau and increase to peak), and the ejection phase (rapidly cooling post-peak). The temperature evolution of material ejected near the disk edge is shown in dotted lines and the material ejected close to the neutron star is shown in solid lines.}
    \label{fig:simple_comp}
\end{figure}

A sample of the trajectories produced in this 3-phase formalism is shown in Fig.~\ref{fig:simple_comp} using our high angular momentum models and an $\alpha=0.01$. The angular momentum of the accreting material produces a long-lived disk phase (over 1000s). This long lived disk evolution phase can be split into two sections. The plateau section is the sub giga-Kelvin temperature region, where material first enters the edge of the accretion disk and begins to lose angular momentum. The peak section is a rapid temperature increase as the material reaches its disk ejection radius relative to the neutron star. As the material moves through the disk it is compressed further due to the gravitational effects of the neutron star, therefore increasing the drag forces and further dissipating the angular momentum, leading to the exponential increase in temperature, until the material is ejected.

For nuclear burning in common-envelopes, the yields appear to be insensitive to the nature of the first (infall) phase of our trajectories. The free-fall phase evolves quickly and the temperatures and densities tend to be sufficiently low that little burning occurs. The yields are most sensitive to the disk evolution stage. At the beginning of the ejection phase the temperature and densities are still sufficiently high for nuclear burning to occur, as the temperature continues to decrease through the ejection phase the material quickly cools and decay processes dominate. 

We assume the neutron star accretion disk ejects material into the common envelope at a constant rate across the disk surface. The final isotopic mass fractions from each ejection radius, for a single accretion rate, were averaged to produce a final isotopic mass fraction distribution for that accretion rate. This averaging provides insight into the composition of the material that is ejected back into the common envelope. To understand how material is being processed in the accretion disk we look at the instantaneous reaction flux and isotopic evolution for a trajectory with an individual point of ejection. 

Table \ref{tab:table1} shows the range of peak temperatures reached in each trajectory, as well as the temperature at the edge of the accretion disk. The plateau temperatures of each accretion rate are above the threshold for CNO nuclear burning. This will provide a mechanism for material to be processed into heavier isotopes before the peak temperature is reached.

\begin{table}\centering
\caption{Details of trajectories studied in the present work. The first column shows the accretion rate that feeds the accretion disk, the second column shows the range of distances from the centre of the neutron star where material is ejected, column three shows the temperature of the material as it orbits in the edge of the accretion disk, and the final column shows the peak temperature range of the material before it is ejected}\label{tab:table1}
\begin{tabular}{l|l|c|c}\toprule
\parbox{1.7cm}{Accretion rate (M$_{\odot}$~s$^{-1}$)} &\parbox{1.7cm}{Range of ejection ($\times10^{6}$ cm)} &\parbox{1.7cm}{Plateau temperature (GK)} &\parbox{1.7cm}{Peak temperature range (GK)} \\\midrule
1$\times10^{-5}$ &1.0422 - 4.9695 &0.094 &0.86 - 2.61 \\
16$\times10^{-5}$ &1.0422 - 4.9695 &0.191 &1.71 - 5.23 \\
32$\times10^{-5}$ &1.0422 - 4.9696 &0.224 &2.04 - 6.21 \\
512$\times10^{-5}$ &2.0355 - 4.9695 &0.454 &4.07 - 8.72 \\
\bottomrule
\end{tabular}
\end{table}

Neutrino emission and absorption can vary the electron fraction in rapidly accreting disk calculations.  However, for the low accretion rates used in this study, the deleptonization from electron capture and any neutrino absorption is minimal. We will discuss this in more detail in a later paper.

\section{Companion star details}
\label{sec:NuGrid_network}

The companion star plays a significant role in both the evolution of the common envelope system and the type of material that is accreted around the neutron star. The stellar evolution of a 15 M$_{\odot}$ zero age main sequence (ZAMS) star with metallicity of 2\% from \citet{2018MNRAS.480..538R} was used to calculate the composition and accretion rate at various points through the expansion phase of the star. Specifically, when the core of the star begins to deplete hydrogen at its centre through the CNO cycle, helium is produced in the core and the star begins to contract. This contraction causes heating, which in turn triggers shell hydrogen burning. The increased energy output from hydrogen shell burning further drives an expansion of the stellar envelope ~\citep{2007nps..book.....I}.  

\begin{figure}
    \centering
    \includegraphics[width=\columnwidth]{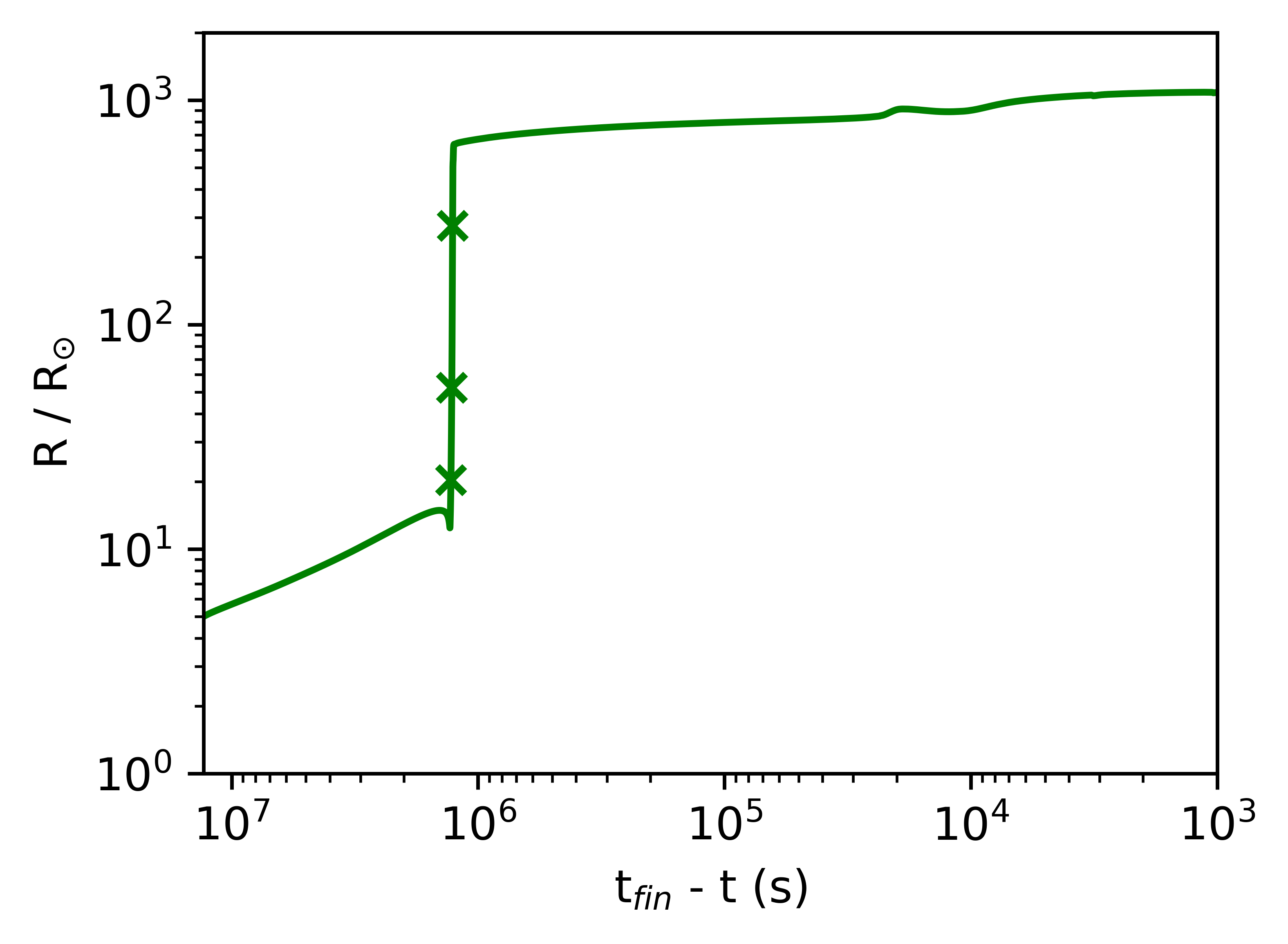}
    \caption{Evolution of the 15 M$_{\odot}$ companion. The x-axis shows the time remaining until core collapse and the y-axis shows the radius of the companion. The markers show the time during the companion evolution at which we modelled the common envelope events. The companion ages used are: $1.256\times10^{7}$, $1.257\times10^{7}$ and $1.258\times10^{7}$ years}
    \label{fig:placeholder}
\end{figure}

The evolution of the 15M$_{\odot}$ companion is shown in Fig. \ref{fig:placeholder}, the three points on the line show the points during expansion where the compositions at specific accretion rates were calculated. As the neutron star accretes material, a velocity and density gradient will form across the accretion disk \citep{2015ApJ...798L..19M}. These gradients will limit the rate at which material accretes. To account for this, the accretion rate calculated for each point in the companion star is multiplied by a non-dimensional parameter $\lambda_{\textrm{BHL}}$. In this work, $\lambda_{\textrm{BHL}} = 0.1$. The inclusion of the parameter $\lambda_{\textrm{BHL}}$ follows \citet{Keegans_2019}, where $\lambda_{\textrm{BHL}}$ had a upper and lower value: 0.25 and 0.025 respectively. 

During this phase of rapid expansion, three companion sizes (ages) were used to calculate the initial composition of the material entering the neutron star accretion disk; 20.3 R$_{\odot}$ ($1.256\times10^{7}$ years), 52.4 R$_{\odot}$ ($1.257\times10^{7}$ years) and 275.6 R$_{\odot}$ ($1.258\times10^{7}$ years). These three radii were chosen as the formation of the common envelope will depend on the orbital separation of the neutron star and its companion. If they are close the common envelope will form early on during the expansion phase, if they are distantly separated it will form at the end. As the neutron star plunges towards the companion core it will accrete material the entire time, however material accreted early on in the plunge in phase is not dense and does not reach high enough temperatures for interesting nucleosynthesis to occur. Once the neutron star is within 5 solar radii the mass transferred is sufficient to produce the high temperature and density medium that result in complex nucleosynthesis. The accretion rates investigated in this study range between $1\times10^{-5} $ M$_{\odot}~$s$^{-1}$ and $512\times10^{-5} $ M$_{\odot}~$s$^{-1}$ which produce peak temperatures between 0.856 GK and 8.72 GK.

\section{Nucleosynthesis calculations and reaction network}

As the neutron star accretion disk is a high temperature and density environment, ranging from 0.25 GK and 1 g~cm$^{-3}$ near the edge of the accretion disk up to 8.72 GK and $3.29\times10^{8}$ g~cm$^{-3}$ near the neutron star surface, a variety of different types of nucleosynthesis can occur. The environment also rapidly changes as the material evolves through the accretion disk and is then ejected. To ensure the large variety of potential reaction pathways and rapid changes in the temperature and density are correctly modelled the NuGrid nucleosynthesis tool PPN\footnote{https://nugrid.github.io/} \citep{2016ApJS..225...24P,jones2019a} was used to post process each trajectory. This utilised an isotopic network spanning from neutrons and protons up to $^{216}$Po including all isotopes with a half-life greater than $5\times10^{-6}$ seconds. This decay limit was chosen as we observed no reaction flux going through the network boundary. This allows short lived isotopes that would decay before the end of the scenario to be identified. This results in a network of 5234 isotopes split across the valley of stability. A large network was used as the high temperature leads to production of exotic nuclei away from stability.

The reaction rate data used in NuGrid comes primarily from the Joint Institute for Nuclear Astrophysics Center for the Evolution of the Elements, JINA-CEE, Reaclib compilation \citep{Cyburt_2010}. By default NuGrid is designed to use JINA-CEE Reaclib v1.1, published in April 2013, which is no longer up-to-date. To ensure correct reaction data was used the JINA-CEE Reaclib default compilation (upadated on the 24th of June 2021), published in June 2021, was used in place of Reaclib v1.1. The Reaclib library does not cover the entire network of reactions and so is supplemented with neutron capture rates from the Karlsruhe Astrophysical Database of Nucleosyntheis in Stars (KADoNiS) \citep{dillmann2006kadonis}, as well as beta decay rates by \citet{ODA1994231} and \citet{LANGANKE2000481}.

\section{Initial mass fraction variations}
\label{sec:iniab}

\begin{figure}
    \centering
    \includegraphics[trim = {0cm 0cm 0cm 0cm}, clip,width=\linewidth]{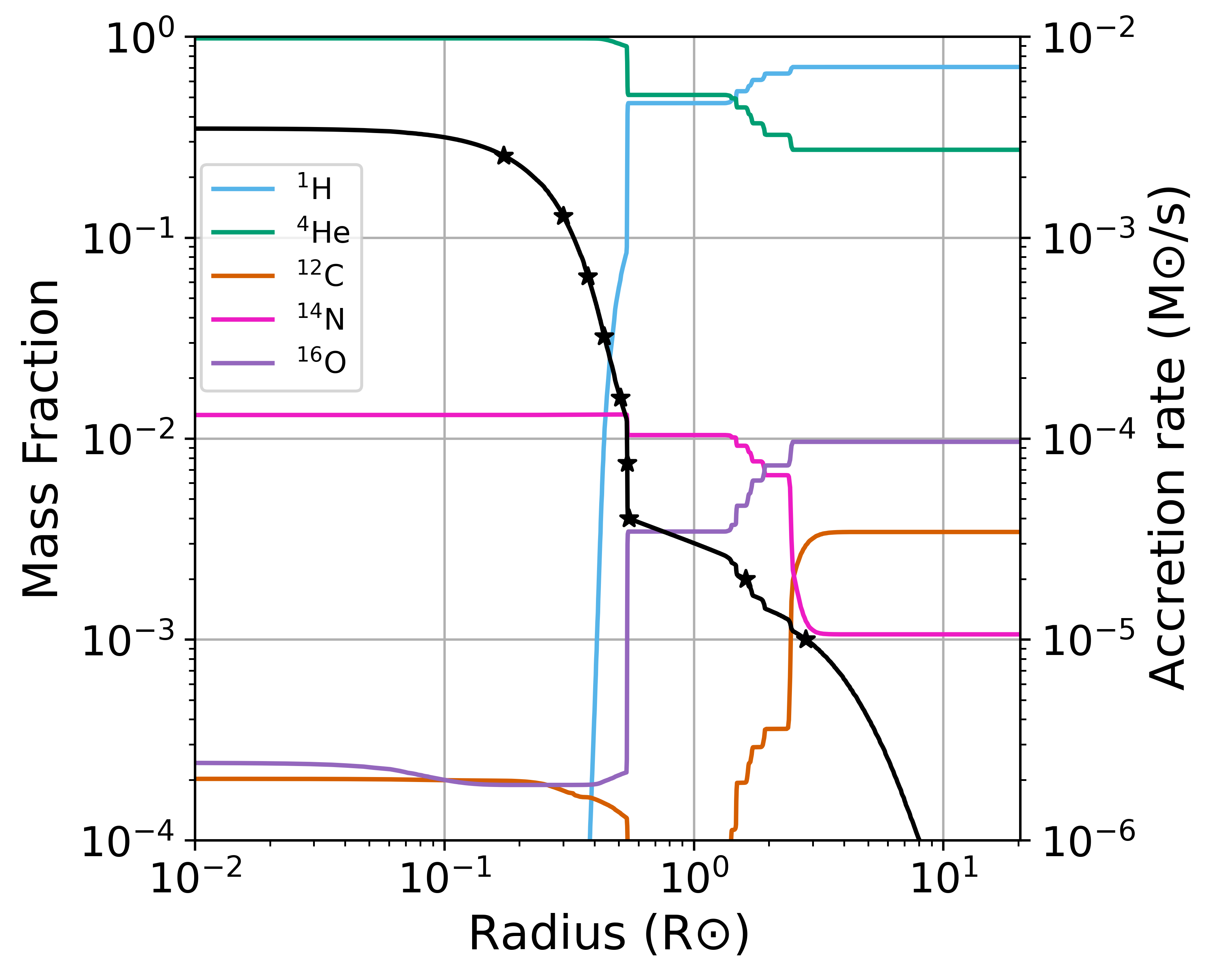}
    \caption{\textbf{}Composition of a 15 M$_{\odot}$ ZAMS star, when it has expanded to 20.3 R$_{\odot}$. The mass fraction (left axis) for $^{1}$H, $^{4}$He, $^{12}$C, $^{14}$N and $^{16}$O shown as a function of radial distance from the companion core. Also shown is the accretion rate (right axis) in solar masses per second vs radial distance. Black asterisks indicate accretion rates used in current study.}
    \label{fig:H_frac_to_iniab_early}
\end{figure}

The position of the neutron star inside the companion envelope will impact the composition of the accreted material. Fig. \ref{fig:H_frac_to_iniab_early} shows how the mass fractions of $^{1}$H, $^{4}$He, $^{12}$C, $^{14}$N and $^{16}$O vary with respect to radial position inside the companion envelope, as well as how the accretion rate varies throughout the 20.3 R$_{\odot}$ 15$M_\odot$ companion after core hydrogen burning has finished but before core helium burning begins. This clearly shows that the highest accretion rates, $32\times10^{-5}$ M$_{\odot}$ s$^{-1}$ to $256\times10^{-5}$ M$_{\odot}$ s$^{-1}$, can only be achieved inside the helium core, therefore the material entering the accretion disk will have the same composition as the companion core.

The initial composition for each accretion rate was calculated using MESA stellar evolution models by \citet{2018MNRAS.480..538R}. Using the evolution of a 15$M_{\odot}$ star, the internal structure and composition were taken at three times post core hydrogen burning. We model only the plunge in phase of the common envelope evolution, as defined by \citet{Ivanova_CE_evolution}, as this is the most studied and well known phase \citep{Passy_2012}. During this phase, the neutron star inspirals through the companion envelope towards the core, transitioning through internal structures of the companion. The inspiral result in the neutron star reaching higher accretion rates and in variation in the composition of accreting material. The initial compositions used contained 170 isotopes between $^{1}$H and $^{62}$Ni taken from MESA while solar seed nuclei up to bismuth are taken from NuGrid stellar yields, both originate from \citet{2018MNRAS.480..538R}. 

\begin{figure}
    \centering
    \includegraphics[trim = {.25cm .25cm 0 .7cm}, clip,width=\linewidth]{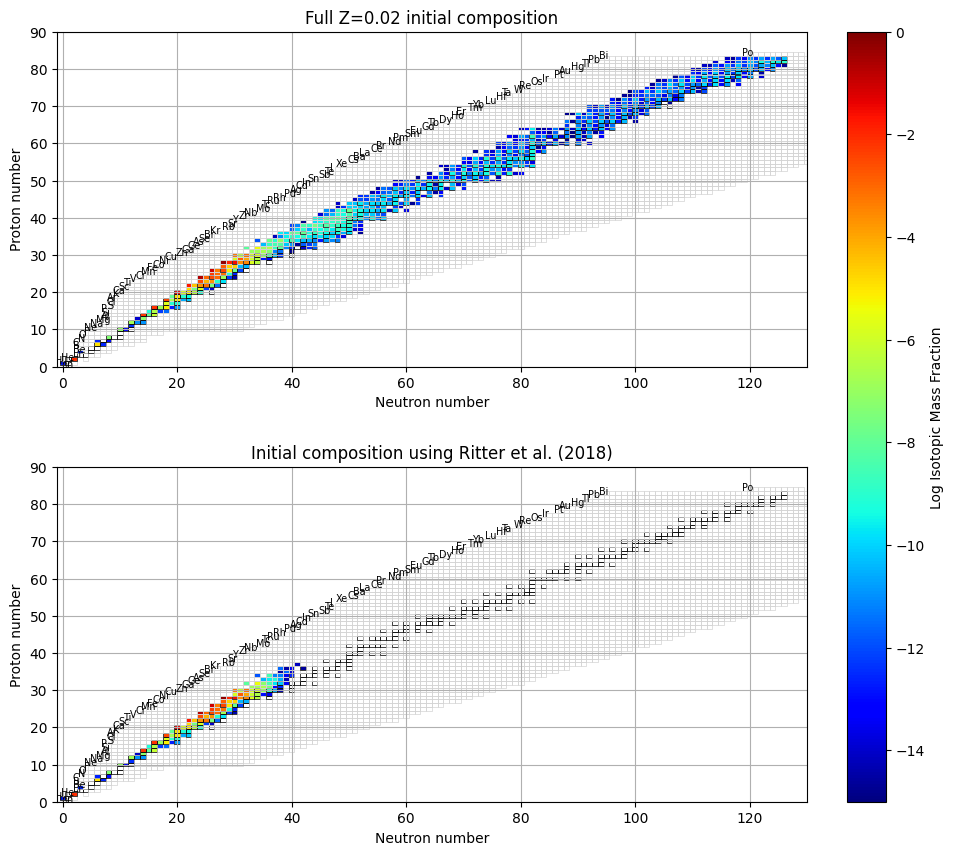}
    \caption{Final isotopic mass fractions for a 1.5 M$_{\odot}$ neutron star accreting 32$\times10^{-5} \textrm{M}_{\odot}\textrm{s}^{-1}$ from a 15 M$_{\odot}$ companion. The top shows the resulting isotopic distribution when using the initial composition from \citet{2018MNRAS.480..538R} and including all solar seed nuclei (389 isotopes between $^{1}\textrm{H}$ and $^{209}\textrm{Bi}$). The bottom figure shows the resulting isotopic distribution when using an initial composition taken from \citet{2018MNRAS.480..538R} MESA models (170 isotopes between $^{1}\textrm{H}$ and $^{62}\textrm{Ni}$).}
    \label{fig:15M_iso_chart_iniab_comp}
\end{figure}

It is important to include the full range of initial seed nuclei. Fig. \ref{fig:15M_iso_chart_iniab_comp} shows the resulting final isotopic distribution when using only 170 isotopes, up to $^{56}$Ni, used from \citet{2018MNRAS.480..538R} MESA models (bottom) compared to the the initial composition containing the same 170 isotopes up to $^{56}$Ni as well as all stable solar seed nuclei between $^{56}$Ni and $^{209}$Bi. Without the inclusion of these heavy seeds, nucleosynthesis only occurs in the low mass region, but with the heavy seeds included there is nucleosynthesis throughout the chart of nuclei.

\section{Nucleosynthesis during common envelope evolution}
\label{sec:results}

All of the data presented in this section utilise a single value for the angular momentum of the accreted material, $1\times10^{17}$ cm$^{2}$~s$^{-1}$, and therefore for a single size of accretion disk. As part of this work, we studied the effect of changing the angular momentum of accreted material. Decreasing the angular momentum of the accreted material by a factor of 10 results in a longer plateau time, therefore producing more CNO and CNO breakout isotopes. However, as the nuclear burning is most sensitive to the peak temperature, the resulting isotopic and elemental distributions still follow the same trends as those in the high angular momentum case.

\subsection{Hydrogen rich outer envelope region}
\label{sec:rp-process_results}

\begin{figure}
    \centering
    \includegraphics[trim = {.25cm .2cm .25cm 1.05cm}, clip,width=\linewidth]{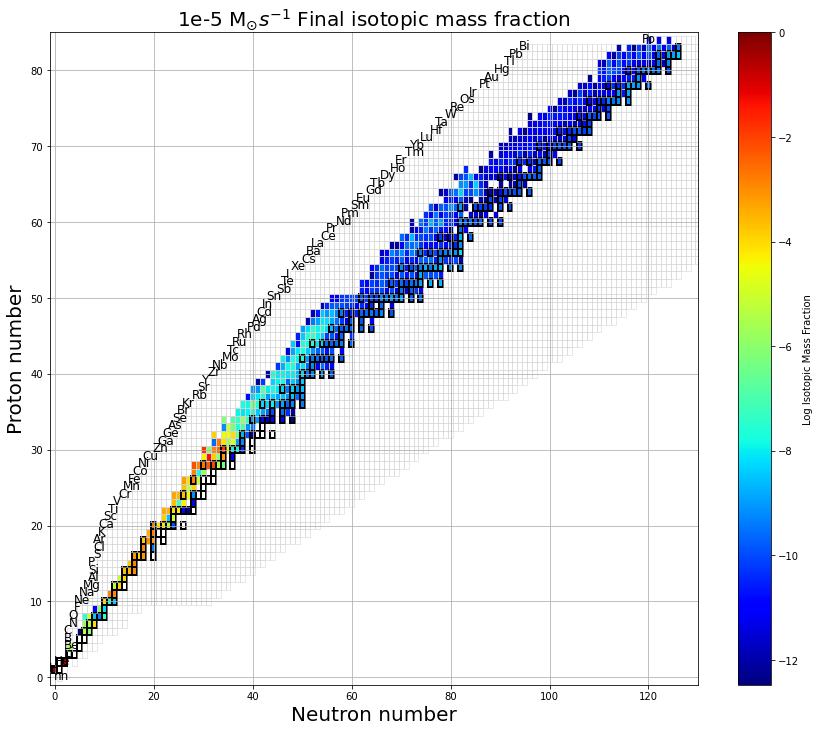}
    \caption{Average undecayed final isotopic distribution from the $1\times10^{-5} \textrm{M}_\odot~\textrm{s}^{-1}$ accretion rate trajectory. Average over disk ejection radius ($1.0422\times10^{6} - 4.9695\times10^{6}$ cm). Stable nuclei shown in black boxes.}
    \label{fig:1e-5_final_iso_chart}
\end{figure}

A common envelope evolves when the companion star first undergoes expansion to a radius of 20.3 R$_{\odot}$. The lowest accretion rate studied is 1$\times10^{-5}$ M$_{\odot}$~s$^{-1}$. Fig. \ref{fig:H_frac_to_iniab_early} shows that the neutron star accretes material from the bottom of the hydrogen burning shell. The initial composition consists of $70.6\%$ hydrogen, with stable nuclei and trace amounts of $^{13}$N, $^{15}$O and $^{22}$Na that are produced during the lifetime of the companion. $^{13}$N and $^{15}$O are produced during the CNO cycle inside the companion, $^{22}$Na is produced through proton captures and has a half-life of 2.6 years. Fig. \ref{fig:1e-5_final_iso_chart} shows the averaged mass fraction of isotopes ejected throughout the accretion disk. The high abundance of protons combined with the high temperature and density through out the trajectory (see Fig. \ref{fig:simple_comp}) results in the synthesis of neutron deficient material. There is evidence of material reaching the SnSbTe cycle, as the mass fractions of isotopes below the SnSbTe cycle have comparatively high mass fractions.

\begin{figure}
    \centering
    \includegraphics[trim = {0cm 0cm 0cm .67cm}, clip,width=\linewidth]{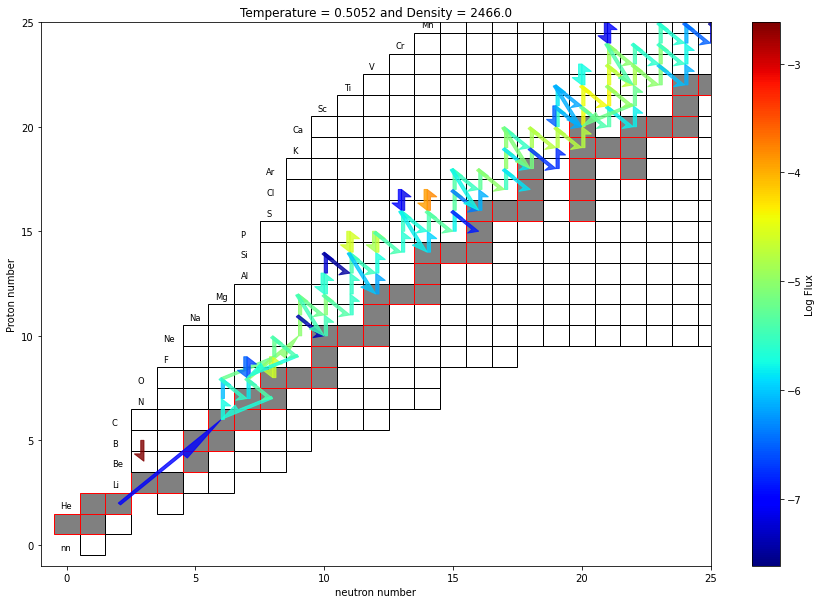}
    \caption{Reaction flux through the $1\times10^{-5}$ M$_{\odot}$~s$^{-1}$ accretion rate trajectory for the $2.0355\times10^6$ cm disk ejection radius at 0.5 GK, before the peak temperature is reached, showing the HCNO cycle breakout to rp-process. The solar-like initial composition was calculated using MESA models from \citet{2018MNRAS.480..538R}. }
    \label{fig:0.5GK_rp-process_HCNO}
\end{figure}

\begin{figure}
    \centering
    \includegraphics[trim = {0cm 0cm 0cm .67cm}, clip,width=\linewidth]{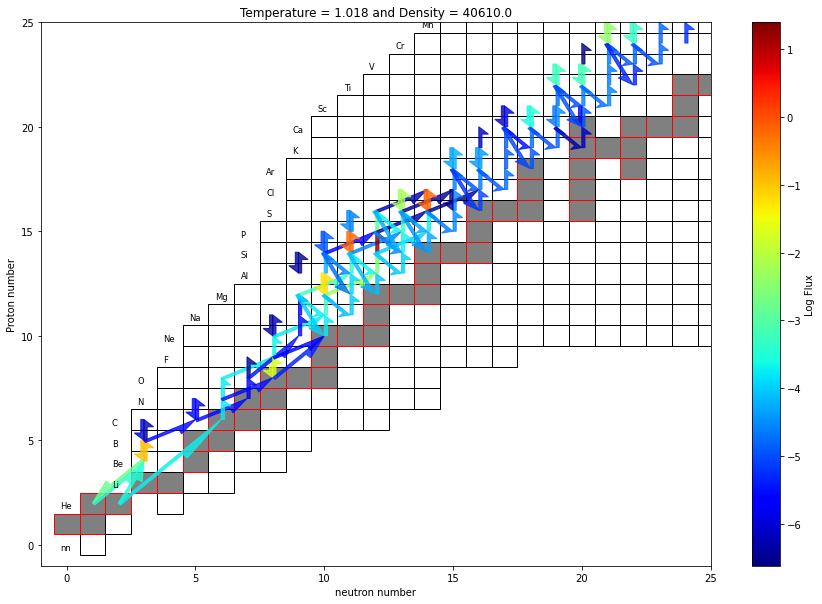}
    \caption{Reaction flux through the $1\times10^{-5}$ M$_{\odot}$~s$^{-1}$ accretion rate trajectory at 1 GK before the peak temperature is reached. }
    \label{fig:1GK_rp-process_breakout}
\end{figure}

\begin{figure}
    \centering
    \includegraphics[trim = {0cm 0cm 0cm .67cm}, clip,width=\linewidth]{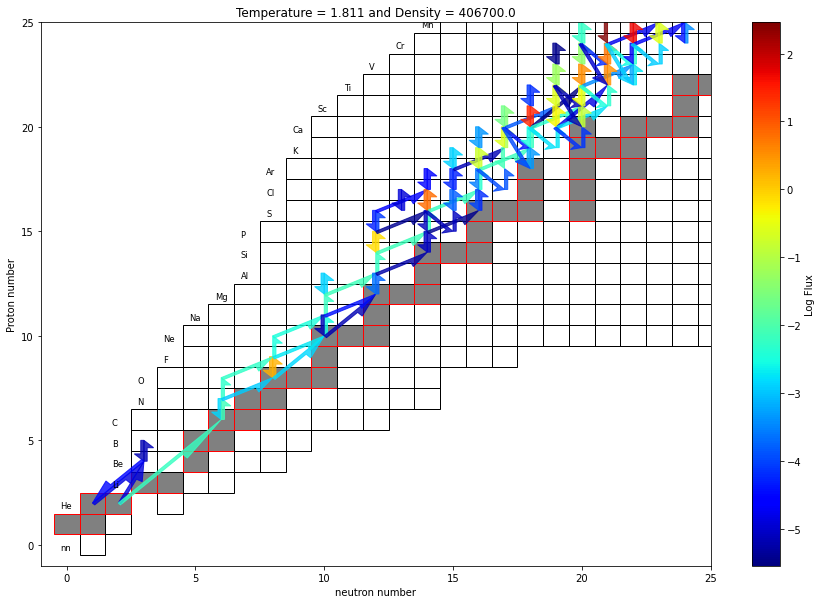}
    \caption{Reaction flux through the $1\times10^{-5}$ M$_{\odot}$~s$^{-1}$ accretion rate trajectory at the peak temperature, 1.8 GK.}
    \label{fig:Peak_T_rp-process_breakout}
\end{figure}

Flux plots are utilised to investigate the reaction flow of material at three points during this trajectory, at temperatures of 0.5GK (Fig. \ref{fig:0.5GK_rp-process_HCNO}), 1GK (Fig. \ref{fig:1GK_rp-process_breakout}) and the peak temperature of the trajectory which is 1.8GK (Fig. \ref{fig:Peak_T_rp-process_breakout}). For the full temperature evolution of this trajectory see Fig. \ref{fig:simple_comp}. 

Fig. \ref{fig:0.5GK_rp-process_HCNO} shows the reaction flux as material reaches 0.5 GK. At this temperature the reaction flux clearly shows the hot CNO cycle and the breakout reaction $^{15}\textrm{O}(\alpha,\gamma)^{19}\textrm{Ne}$ have high reaction fluxes. Proton capture chains do not reach the proton drip line, as shown by the balanced proton capture and proton emission reactions at $^{25}$Si, $^{30}$S and $^{42}$Ti. Proton captures reach up to $^{56}$Fe, but cannot process material further due to competing beta decay reactions. 

Fig. \ref{fig:1GK_rp-process_breakout} shows the reaction flux at 1GK, before the peak temperature is reached. We see that the main nucleosynthesis pathways have changed when compared to Fig. \ref{fig:0.5GK_rp-process_HCNO}. At this higher temperature, $\alpha$-capture reactions begin to have a significant role in producing heavier isotopes. The $^{18}\textrm{Ne}(\alpha,p)^{21}\textrm{Na}$, $^{22}\textrm{Mg}(\alpha,p)^{25}\textrm{Al}$ and $^{26}\textrm{Si}(p,\gamma)^{27}\textrm{P}$ reactions provide a pathway for material to escape the hot CNO cycle and enter the rp-process. Between magnesium and chlorine, the reaction channels compete between alpha capture and proton captures. Beyond chlorine, proton capture reactions are the main source of nucleosynthesis. When compared to Fig. \ref{fig:0.5GK_rp-process_HCNO}, we see that material is reaching further into the neutron deficient region and the competing proton capture and beta decay channels are further from the valley of stability. At this temperature proton capture reactions extend past the iron group up to gallium and germanium, but do not extend to the SnSbTe cycle isotopes.  

Fig. \ref{fig:Peak_T_rp-process_breakout} shows the reaction flux at the peak temperature, 1.8GK. The extreme temperature and presence of both helium and hydrogen result in a combination of $(\alpha,p)$, $(\alpha,\gamma)$ and $(p,\gamma)$ which drive nucleosynthesis up to phosphorus. Beyond phosphorus, nucleosynthesis is dominated by proton capture reactions. Proton captures reach the SnSbTe cycle isotopes at the peak temperature, as shown in Fig. \ref{fig:1e-5_final_iso_chart}.

\begin{figure}
    \centering
    \includegraphics[width=\linewidth]{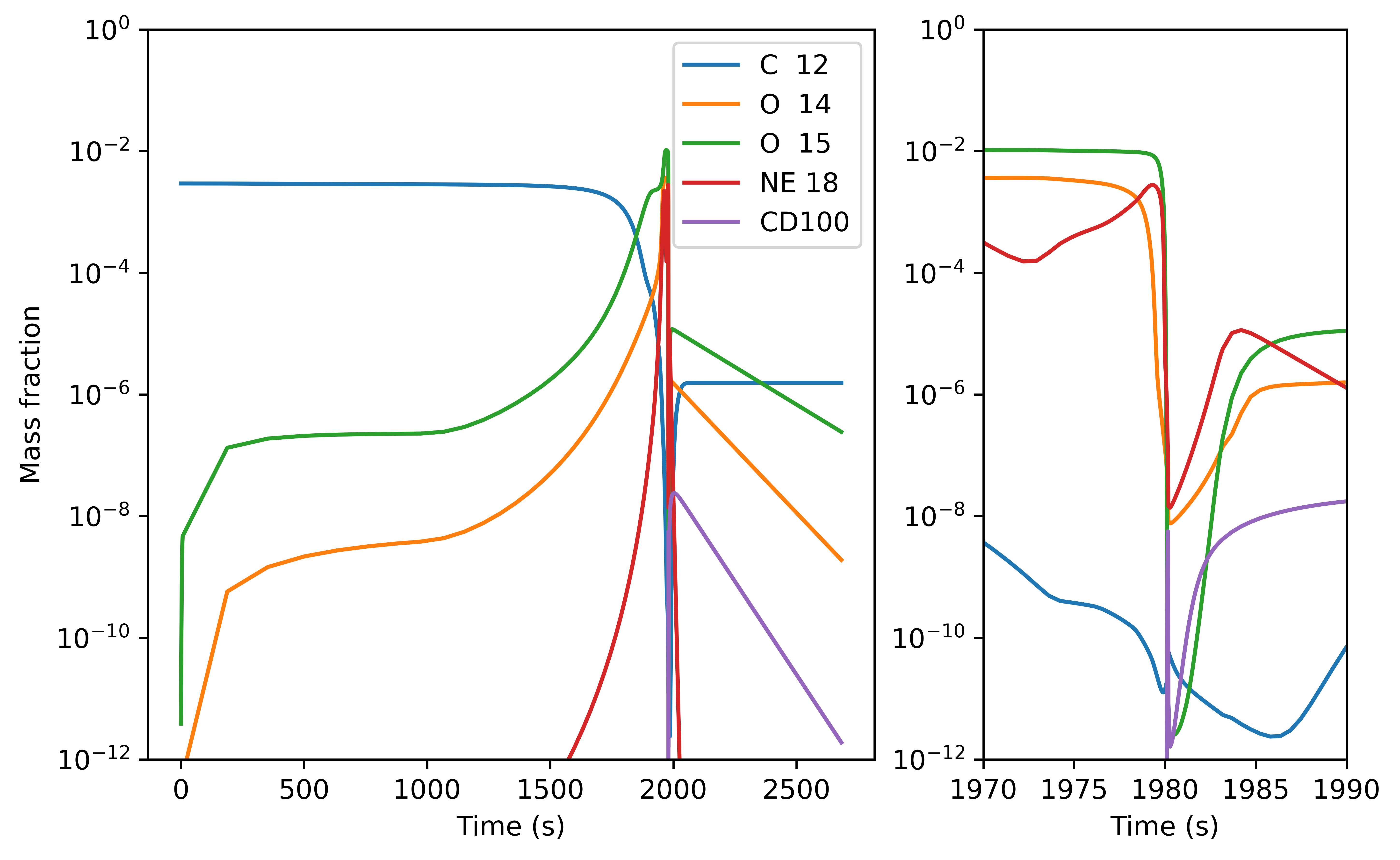}
    \caption{Isotopic evolution of isotopes around CNO during $1\times10^{-5}$ M$_{\odot}$~s$^{-1}$ accretion rate trajectory. $^{100}$Cd is included showing that the rp-process reaches the SnSbTe cycle.}
    \label{fig:CNO_iso_evo_rp-process_1e-5}
\end{figure}

Fig. \ref{fig:CNO_iso_evo_rp-process_1e-5} shows the isotopic evolution of $^{12}$C, $^{14,15}$O, $^{18}$Ne and $^{100}$Cd for the entire length of the trajectory (left), and zoomed in around the peak temperature (right). $^{12}$C is slowly destroyed and $^{14}$O and $^{15}$O are produced via the CNO cycle during the lead up to the peak temperature. At the peak temperature, these isotopes are destroyed via proton and alpha capture reactions. Due to the high availability of protons and high temperature after material has reached its disk ejection radius (peak temperature) relative to the neutron star, $^{12}$C and $^{15}$O are produced again as the CNO cycle is able to continue processing material during the disk ejection phase. $^{100}$Cd is produced during the peak temperature of the trajectory. The sharp peak at 1980 seconds shows $^{100}$Cd being produced and then immediately destroyed as proton captures process material through $^{100}$Cd. Shortly after the peak temperature (approximately 5 seconds after) of the trajectory, the mass fraction of $^{100}\textrm{Cd}$ increases as a result of $\beta^{+}$ decays on the isobar (right panel of Fig. \ref{fig:CNO_iso_evo_rp-process_1e-5}). This indicates that nucleosynthesis of reached the SnSbTe cycle before it is ejected. As $^{100}\textrm{Cd}$ is radioactive, it then decays to stability as material is ejected from the neutron star accretion disk.

\begin{figure}
    \centering
    \includegraphics[width=\linewidth]{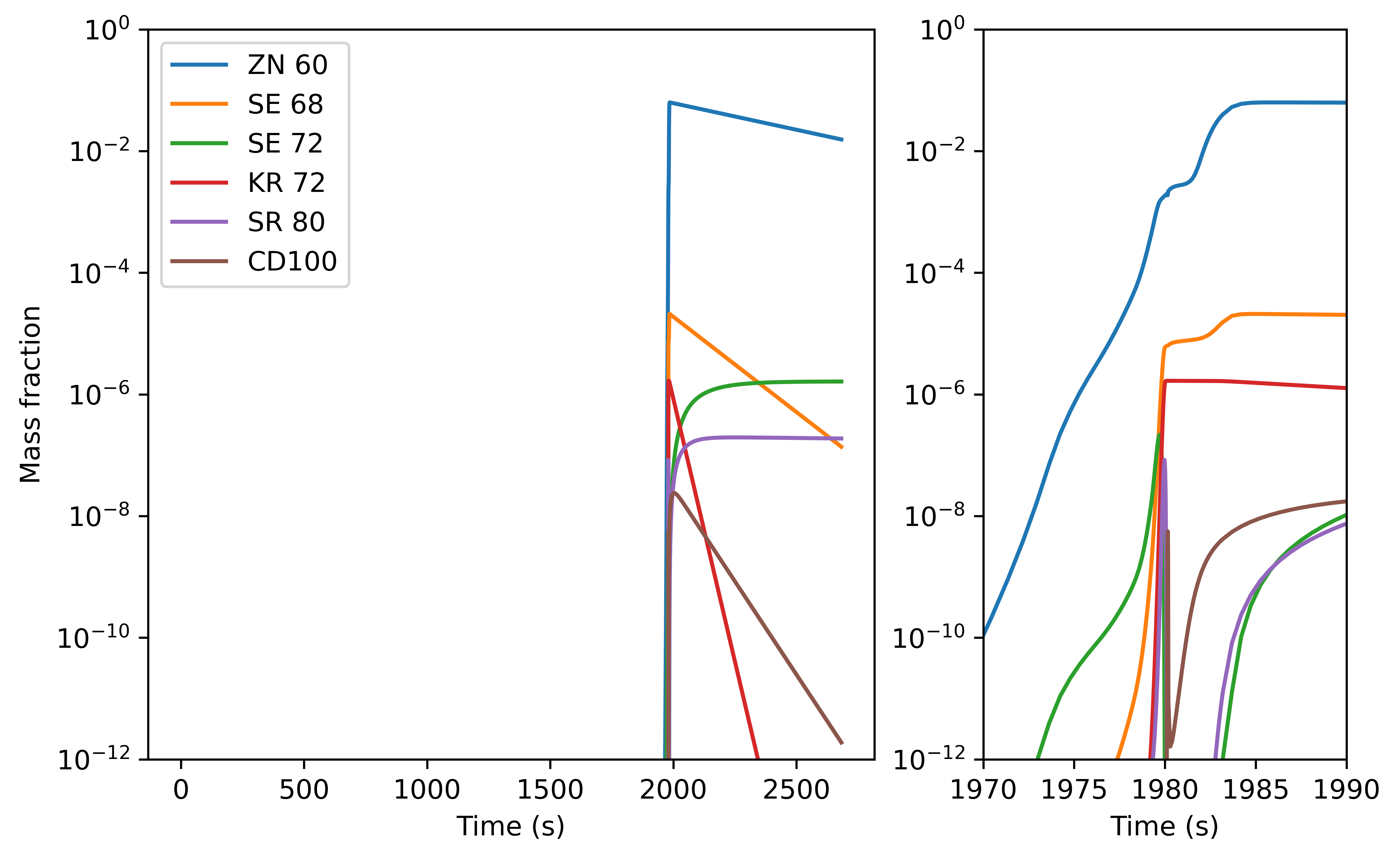}
    \caption{Isotopic evolution of heavy rp-process waiting point isotopes during $1\times10^{-5}$ M$_{\odot}$~s$^{-1}$ accretion rate trajectory. $^{100}$Cd is included as this shows the rp-process reaching the SnSbTe cycle.}
    \label{fig:Heavy_iso_evo_rp-process_1e-5}
\end{figure}

Fig. \ref{fig:Heavy_iso_evo_rp-process_1e-5} provides further evidence that the high mass fraction of $^{100}\textrm{Cd}$ in Fig. \ref{fig:CNO_iso_evo_rp-process_1e-5} and high undecayed final mass fraction of isotopes below the SnSbTe cycle in Fig. \ref{fig:1e-5_final_iso_chart} are as a result of rp-process nucleosynthesis. $^{60}\textrm{Zn}$, $^{68}\textrm{Se}$, $^{72}\textrm{Kr}$ and $^{80}\textrm{Sr}$ are rp-process waiting point isotopes and material accumulates at these isotopes during rp-process nucleosynthesis. The right panel of Fig. \ref{fig:Heavy_iso_evo_rp-process_1e-5} shows the rapid production and destruction of $^{72}$Se, $^{80}$Sr and $^{100}$Cd at the peak temperature as material is processed through these isotopes towards the proton drip line. it also shows that the mass fraction of $^{60}\textrm{Zn}$ increases followed by that of $^{68}\textrm{Se}$, followed by $^{72}\textrm{Kr}$ as the peak temperature of the trajectory is reached. This is evidence of material moving through these isotopes during the rp-process, leading to production of $^{100}\textrm{Cd}$. Overproduction of p-nuclides is observed in these trajectories, as explored by Abrahams et al. (in prep.).

\subsection{Mixed hydrogen and helium region}
\label{sec:mixed_results}

\begin{figure}
    \centering
    \includegraphics[trim = {0cm 0cm 0cm 1.03cm}, clip,width=\linewidth]{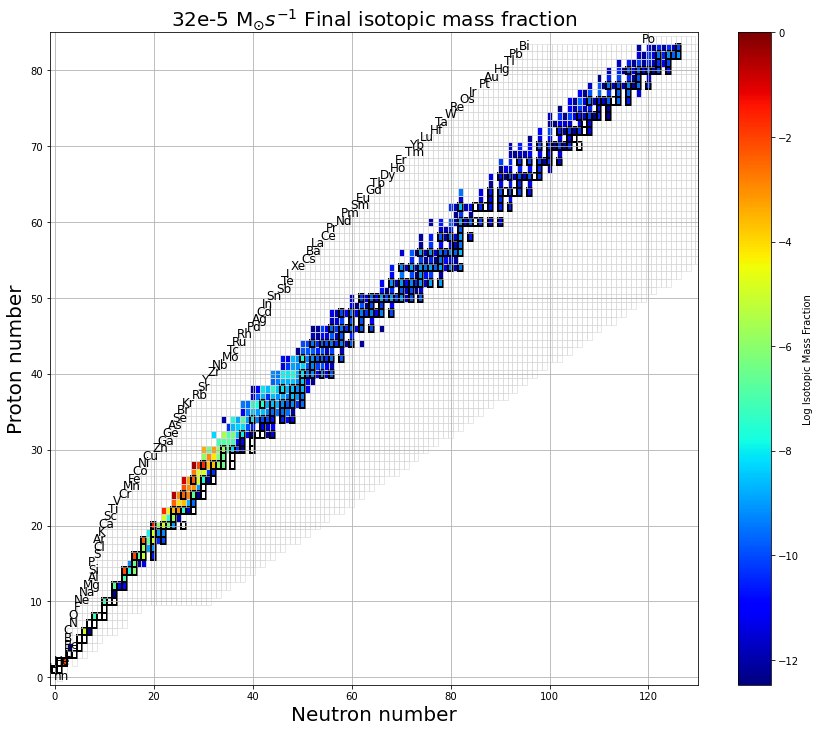}
    \caption{Average final isotopic distributions from the $32\times10^{-5} \textrm{M}_\odot~\textrm{s}^{-1}$ accretion rate trajectory. }
    \label{fig:50/50_final_MF}
\end{figure}

When the neutron star has inspiralled further into the envelope of the 20.3R$_{\odot}$ early expansion phase companion, it begins to accrete material from the mixed boundary between the hydrogen shell and helium core. The neutron star accretes material at $32\times10^{-5} \textrm{M}_\odot~\textrm{s}^{-1}$ (for the composition of the companion star at this accretion rate see Fig. \ref{fig:H_frac_to_iniab_early}). Trajectories with this accretion rate have a peak temperature ranging between 2 GK for material ejected near the edge of the disk and 6.21 GK for material ejected near the neutron star (see Fig. \ref{fig:simple_comp}). The final isotopic abundance for material ejected from the accretion disk at this point in the companion star is shown in Fig. \ref{fig:50/50_final_MF}. When compared to the averaged mass fractions ejected from the $1\times10^{-5}$ M$_{\odot}$ s$^{-1}$, with peak temperatures ranging from 0.856 GK to 2.61 GK (see Fig. \ref{fig:1e-5_final_iso_chart}), we can see that neutron deficient nuclei are produced, but not to the same extent. The bulk of material is trapped in the iron peak elements and nucleosynthesis does not reach the SnSbTe cycle. The final mass fraction of alpha capture products such as $^{28}$Si, $^{32}$S, $^{36}$Ar, $^{40}$Ca and $^{44}$Ti are much higher than the surrounding isotopes, indicating alpha capture reactions drive nucleosynthesis up to the iron group elements.

\begin{figure}
    \centering
    \includegraphics[trim = {0cm 0cm 0cm .68cm}, clip,width=\linewidth]{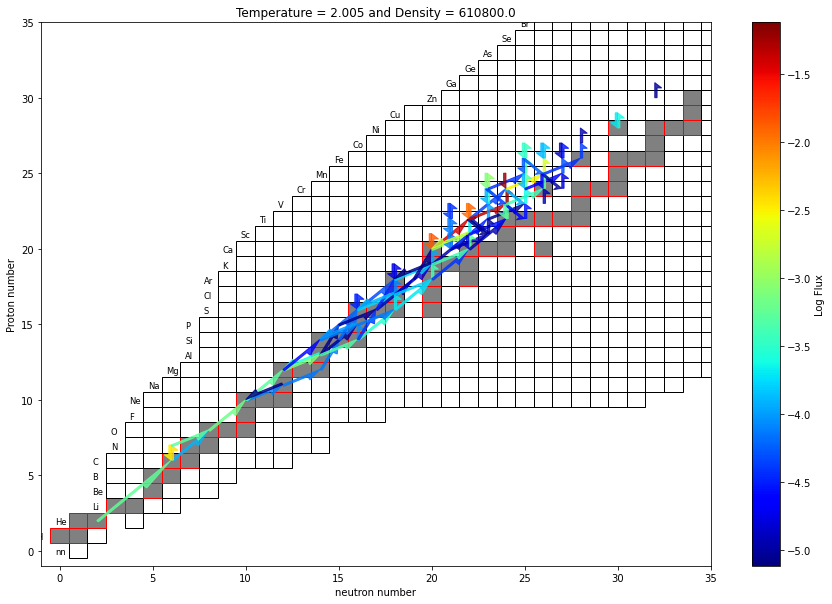}
    \caption{Reaction flux through the $32\times10^{-5}$ M$_{\odot}$~s$^{-1}$ accretion rate trajectory at 2 GK before the peak temperature is reached. This clearly shows the alpha capture process up to $^{44}$Ti. At this temperature the CNO cycle is bypassed by alpha capture reactions. }
    \label{fig:32_peakT_flux}
\end{figure}

Fig. \ref{fig:32_peakT_flux} shows the reaction flux through material accreted during the $32\times10^{-5} \textrm{M}_\odot~\textrm{s}^{-1}$ trajectory when the temperature evolution reaches 2 GK. Alpha captures dominate in the lower mass region, due to the high abundance of alpha particles and the extreme temperatures, which result in successive alpha captures from $^{12}\textrm{C}$ up the iron group, leading to the high final mass fractions of $^{28}$Si, $^{32}$S, $^{36}$Ar, $^{40}$Ca and $^{44}$Ti. Beyond titanium, before the iron group elements, proton capture reactions begin to compete with alpha captures and become the preferred reaction pathway. Beyond iron nucleosynthesis is dominated by proton capture reactions, producing neutron deficient isotopes up to molybdenum. However, due to the reduced availability of protons nucleosynthesis is limited in this region and does not reach up to the SnSbTe cycle.

\begin{figure}
    \centering
    \includegraphics[width=\linewidth]{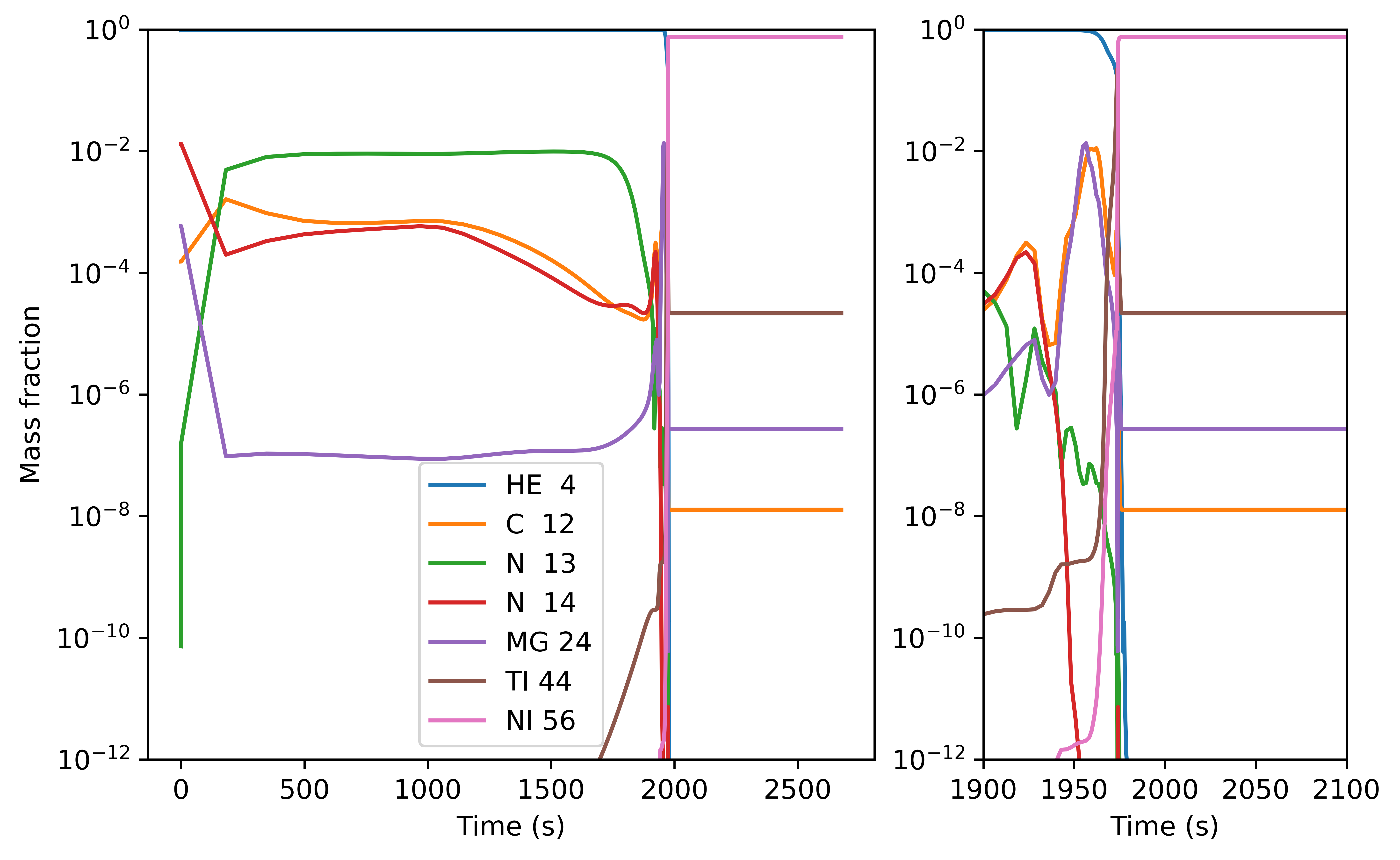}
    \caption{Isotopic evolution of alpha capture products ejected at $2.0355\times10^6$ cm from the centre of the accretion disk during the $32\times10^{-5}$ M$_{\odot}$ s$^{-1}$ accretion rate trajectory. }
    \label{fig:32_alpha_waiting_point_evo}
\end{figure}

To better understand the nucleosynthesis occurring inside the disk, the isotopic evolution of alpha capture isotopes ejected during the $32\times10^{-5}$ M$_{\odot}$~s$^{-1}$ accretion rate was analysed. Fig. \ref{fig:32_alpha_waiting_point_evo} shows the evolution of several alpha capture products, as well as $^{13,14}$N. At the start of the trajectory, during the plateau phase when the temperature is 0.25GK, the CNO cycle is the main source of nucleosynthesis. This is shown in Fig. \ref{fig:32_alpha_waiting_point_evo} by the rapid increase of $^{13}$N early in the time evolution. As the temperature climbs, from 0.25 GK to 4.3 GK at the peak, the fraction of $^{13}$N drops by several orders of magnitude as the CNO cycle is no longer the preferred reaction pathway. As $^{12}$C drops the mass fraction of $^{24}$Mg begins to increase, indicating that alpha capture reactions are bypassing the CNO cycle. This is supported by the rise in the $^{44}$Ti mass fraction as the temperature rises, indicating the alpha capture process is moving material up towards the iron group. Following the peak temperature, $^{4}\textrm{He}$ has been completed depleted and $^{56}$Ni has the highest mass fraction of all isotopes. As we move down the alpha capture products, there is less mass in each isotope. 

\begin{table}\centering
\caption{Mass fraction of $^{44}$Ti and $^{56}$Ni ejected at each point throughout the accretion disk at the end of the $32\times10^{-5} \textrm{M}_\odot~\textrm{s}^{-1}$ accretion rate trajectory.}\label{tab:44Ti_table}
%\resizebox{\textwidth}{!}{ % use this if the table is too large
\begin{tabular}{c|c|c|c|c}\toprule
\parbox{1.2cm}{Disk ejection radius ($\times10^{6}$ cm)} &\parbox{1.2cm}{Mass fraction of $^{44}$Ti} &\parbox{1.2cm}{Mass fraction of $^{56}$Ni}&\parbox{1.2cm}{Peak temperature (GK)} &\parbox{1.2cm}{Ejection velocity (m~s$^{-1}$)} \\\midrule
1.04 &$2.09\times10^{-7}$ &$9.02\times10^{-1}$ &6.21 &$1.96\times10^{8}$ \\
1.30 &$3.33\times10^{-7}$ &$9.15\times10^{-1}$ &5.97 &$1.75\times10^{8}$ \\
1.63 &$5.32\times10^{-7}$ &$9.32\times10^{-1}$ &5.47 &$1.56\times10^{8}$ \\
2.04 &$2.16\times10^{-5}$ &$7.48\times10^{-1}$ &4.36 &$1.40\times10^{8}$ \\
2.54 &$2.37\times10^{-4}$ &$2.56\times10^{-2}$ &3.15 &$1.25\times10^{8}$ \\
3.11 &$1.13\times10^{-2}$ &$1.05\times10^{-2}$ &2.58 &$1.13\times10^{8}$ \\
3.98 &$2.26\times10^{-2}$ &$6.08\times10^{-3}$ &2.27 &$1.00\times10^{8}$ \\
4.97 &$1.55\times10^{-1}$ &$2.30\times10^{-4}$ &2.04 &$8.95\times10^{7}$ \\
\bottomrule
\end{tabular}
\end{table}

Table \ref{tab:44Ti_table} shows that there is a strong inverse relationship between the production of $^{44}$Ti and $^{56}$Ni when the neutron star accretes material at this accretion rate. For material with a large disk ejection radius, where the peak temperature reached is lower, alpha capture chains produce high fractions of $^{44}$Ti but cannot produce significant amounts of $^{56}$Ni. Comparatively, material which has a small disk ejection radius, where higher peak temperatures occur, undergoes alpha captures up to $^{56}$Ni, burning $^{44}$Ti to do so. Of the total amount of $^{44}$Ti ejected from the accretion disk at this accretion rate, 99.9\% is ejected at in the three outermost ejection radii while less than 1\% of the total $^{56}$Ni is ejected at these radii. The inner five ejection radii containing 99.5\% of the total $^{56}$Ni mass fraction, but less than $1\%$ of the ${44}\textrm{Ti}$ mass fraction.There is no disk ejection radius which ejects significant amounts of both $^{44}$Ti and $^{56}$Ni. $^{44}$Ti and $^{56}$Ni are both known gamma-ray observable isotopes \citep{2020ApJ...890...35A}. 

This has implications for potential observations of accreting neutron star common envelope remnants. The mechanism by which the stellar envelope is ejected at the end of the common envelope phase remains an active area of investigation. However, outflows driven by the neutron star accretion disk may contribute to the ejection of the shared envelope. Table \ref{tab:44Ti_table} shows the ejection velocity of material leaving the disk. This, in context of the inverse relationship between $^{44}$Ti and $^{56}$Ni, indicates that the $^{56}$Ni will be ejected faster than the $^{44}$Ti.  This may lead to a common envelope remnant with a high concentration of $^{44}$Ti near the centre, and a high concentration of $^{56}$Ni further out.  If these outflows power supernova-like outbursts, one way to distinguish them would to compare the $^{56}$ and $^{44}$ line profiles (a.k.a. the velocity distribution of the lines).

\subsection{Helium core region}
\label{sec:helium_results}

\begin{figure}
    \centering
    \includegraphics[trim = {0cm 0cm 0cm 1.05cm}, clip,width=\linewidth]{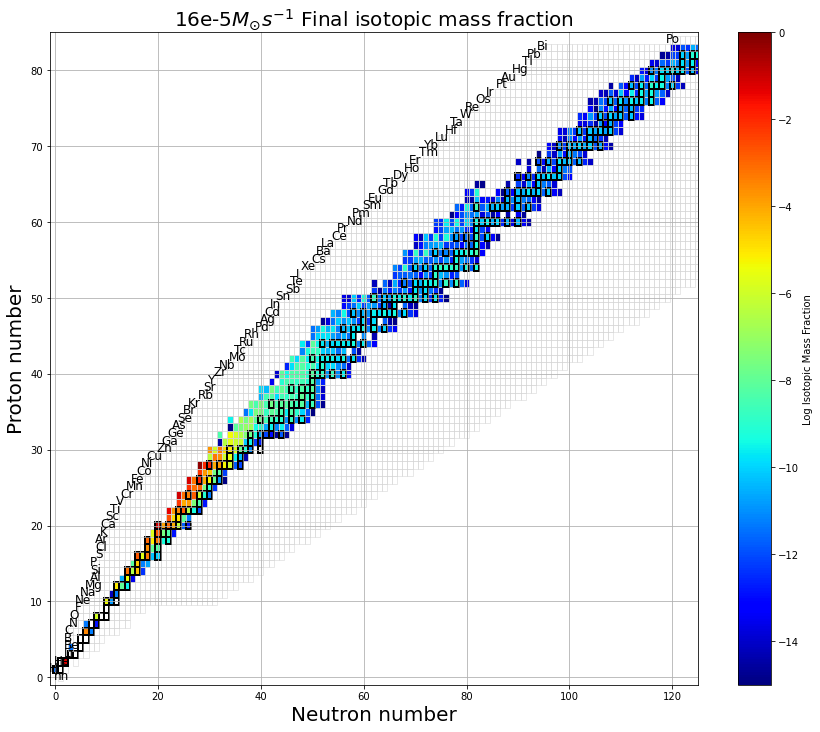}
    \caption{Final isotopic mass fractions averaged over all points of ejection for the $16\times10^{-5}$ M$_{\odot}$ s$^{-1}$ accretion rate inside the 52.4 R$_\odot$ companion.}
    \label{fig:16e-5_midphase_iso_chart}
\end{figure}

As the neutron star inspirals further into the core of the companion star, it begins accreting helium rich material. The initial composition of this environment consists of 93.4\% $^{4}$He and only 0.45\% $^{1}$H by mass fraction. For all three points in the $15\textrm{M}_{\odot}$ star's life explored in this work, the neutron star accretes material from the helium core at an accretion rate between $4\times10^{-5} - 512\times10^{-5}$ M$_{\odot}$~s$^{-1}$ (see Figs \ref{fig:H_frac_to_iniab_early}, \ref{fig:H_frac_to_iniab_mid} and \ref{fig:H_frac_to_iniab_late}). The composition of the material accreted varies between the three companion stars: the $20.3\textrm{R}_{\odot}$ companion has mass fractions of $^{12}\textrm{C}$ on the order of $10^{-4}$, as the star continues to evolve to reach stellar radii of $52.4\textrm{R}_{\odot}$  and then the $275.6\textrm{R}_{\odot}$, the mass fraction of $^{12}\textrm{C}$ increases from $10^{-4}$ to $10^{-3}$. The larger fraction of $^{12}\textrm{C}$ provides a higher fraction of seed nuclei for alpha capture reactions, resulting in higher fractions of alpha capture products. When the neutron star accretes material at $16\times10^{-5}$ M$_{\odot}$ s$^{-1}$, there is evidence of neutron capture reactions. Fig. \ref{fig:16e-5_midphase_iso_chart} shows the final isotopic mass fraction distribution averaged over all ejection points across the accretion disk for the $16\times10^{-5}$ M$_{\odot}$~s$^{-1}$ accretion rate inside the 52.4 R$_\odot$ companion (Fig. \ref{fig:H_frac_to_iniab_mid}). The high fraction of helium results in alpha capture reactions dominating nucleosynthesis, similar in Fig. \ref{fig:32_peakT_flux}. However, there is also a high enough flux of neutrons produced that neutron capture reactions occur and produce trace mass fractions of isotopes on the neutron rich side of the valley of stability. This high neutron flux originates in a combination of $(\alpha,n)$ reactions, which compete with $(\alpha,p)$ and $(\alpha,\gamma)$ reactions in the low mass region and, to a lesser extent, neutron emissions from heavy isotopes due to photodisintegration reactions at high temperatures. 

\begin{figure}
    \centering
    \includegraphics[trim = {0cm 0cm 0cm 1.05cm}, clip,width=\linewidth]{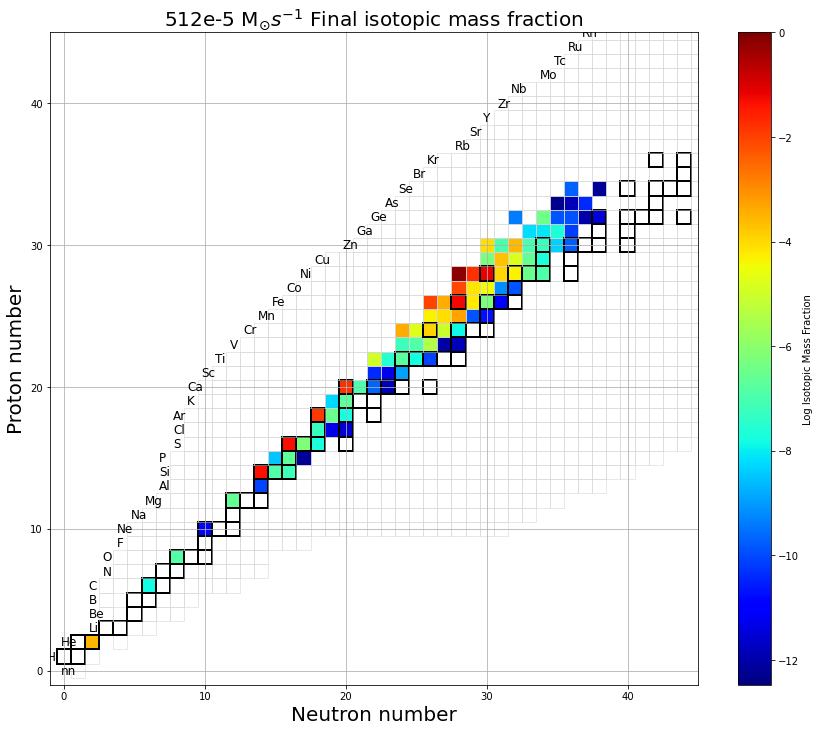}
    \caption{Average undecayed final isotopic distribution for the $512\times10^{-5}$ M$_{\odot}$ s$^{-1}$ accretion rate trajectory. Peak temperatures reached in this trajectory range between 4.07 GK and 8.72 GK. }
    \label{fig:He_core_final_MF}
\end{figure}

The highest accretion rate the neutron star accretes material at is $512\times10^{-5}$ M$_{\odot}$~s$^{-1}$ when the companion has expanded to 52.4 R$_{\odot}$ (see Fig \ref{fig:H_frac_to_iniab_mid}). Trajectories at this accretion rate reach temperatures and densities well into the expected range for nuclear statistical equilibrium. The resulting isotopic distribution for material accreted into the neutron star accretion disk at $512\times10^{-5}$ M$_{\odot}$~s$^{-1}$ is shown in Fig. \ref{fig:He_core_final_MF}. Nucleosynthesis does not reach significantly beyond the iron group elements, and $^{56}$Ni has the highest mass fraction. At these extreme temperatures, the reaction pathways that produce the most stably bound nuclei are favoured. This results in high mass fractions of iron group elements. All accretion rates that exceed $100\times10^{-5}$ M$_{\odot}$~s$^{-1}$ produce iron group elements. This is due to the composition of the accreted material and the extreme temperatures inside the neutron star accretion disk reached as the neutron star inspirals into the helium core. Alpha capture chains produce material up to $^{56}$Ni but cannot continue further. When the neutron star accretes material at $512\times10^{-5}$ M$_{\odot}$~s$^{-1}$, the conditions inside the disk for many of the ejection radii are within the expected NSE temperature range, regardless of the companion star's age.

\section{Conclusions}
\label{sec:conclusion}

From the results shown in section \ref{sec:iniab},
we can see that the full initial composition is essential to modelling nucleosynthesis in the neutron star accretion disk during common envelope. If the full range of seed nuclei are not included then the nucleosynthesis observed is limited, as seen in Fig.\ref{fig:15M_iso_chart_iniab_comp}, and does not accurately represent the nucleosynthetic processes that occur as material moves through the neutron star accretion disk. 

The results of this investigation show that the neutron star accretion disk within a common envelope can produce a variety of types of nucleosynthesis. If material is accreted from the hydrogen rich outer shells, then material in the neutron star accretion disk will undergo rapid proton captures due to the high availability of free protons and the giga-Kelvin temperatures reached during accretion (see Figs \ref{fig:1e-5_final_iso_chart}, \ref{fig:0.5GK_rp-process_HCNO}, \ref{fig:1GK_rp-process_breakout} and \ref{fig:Peak_T_rp-process_breakout}). If the neutron star inspirals to the point that it begins accreting a mix of hydrogen and helium, alpha capture reactions dominate the low mass region up to the iron group. The iron group products then capture protons to produce neutron deficient intermediate mass elements (see Figs \ref{fig:50/50_final_MF}, \ref{fig:32_peakT_flux}). Nucleosynthesis does not reach the end point of the rp-process. When the neutron star accretes helium rich material from the centre of the companion star, it can do so at a range of accretion rates. If the neutron star is towards the outer edge of the helium core then nucleosynthesis can produce trace amounts of neutron rich isotopes close to stability (see Fig \ref{fig:16e-5_midphase_iso_chart}). If the neutron star reaches the centre of the helium core, then the material in the accretion disk experiences extreme temperatures (greater than 5 GK) for enough time to undergo nuclear statistical equilibrium. At these temperatures, the material form iron group elements as these have the highest nuclear binding energy (see Fig.\ref{fig:He_core_final_MF}).

Overall, a variety of nucleosynthetic processes can occur within a neutron star accretion disk within a common envelope, with their products subsequently mixed back into the companion star via wind-driven ejection. At the end of common envelope evolution, this material may be ejected into the interstellar medium and contribute to galactic chemical evolution. 

The inclusion of angular momentum is essential in accurately modelling the neutron star accretion disk within the common envelope. When the angular momentum of the accreting material is not included, such as the work conducted by \citet{Keegans_2019}, the timescale for accretion and ejection is much shorter and so the time in which the material can undergo nucleosynthesis is extremely short. When the angular momentum is included, the material must shed this before it can move further through the accretion disk. This results in the long disk evolution stage and the temperature plateau seen in Fig.\ref{fig:simple_comp}. The disk evolution stage provides a site for CNO and CNO breakout reactions to occur, producing higher fractions of isotopes below the iron peak and providing a larger fraction of seed nuclei to undergo nucleosynthesis at the peak temperature of the trajectory. In particular, the long disk evolution stage allows for seed nuclei in the accreted material to undergo the rp-process. This leads to the ejection of neutron deficient isotopes from the neutron star accretion disk, with potential for them to be ejected into the interstellar medium and contribute to galactic chemical evolution. We found decreasing the angular momentum of the material accreted increases the time spent in the outer region of the accretion disk, and so varies the time spent at the plateau temperature. As the isotopic yields are most sensitive to the peak temperature this results in variation in the CNO and CNO break out isotopes produced during the plateau temperature, but the final mass fraction variations follow the same trends seen in the high angular momentum case. 

We find that $^{44}\textrm{Ti}$ is produced in the outer regions of the accretion disk while $^{56}\textrm{Ni}$ is produced in the inner regions of the disk, and that the ejection velocity of these isotopes varies (see Table. \ref{tab:44Ti_table}). If common envelope events produce supernova-like outbursts, the differences in the abundances may be the primary way to distinguish these events from other supernova engines. For example, since $^{56}$Ni is primarily produced in the inner disk where ejection velocities are higher whereas $^{44}$Ti is produced further out, we would expect the iron (which is the decay product of $^{56}\textrm{Ni}$) in this type of transient remnant to be further out than the $^{44}$Ti. This is currently seen in the Cassiopeia A remnant, but note that there are other aspects of Cassiopeia A that suggest a convective engine~\citep{2014Natur.506..339G}. Measuring the velocity distribution in the $^{44}$Ti lines in Supernova 1987A and comparing it to the $^{56}$Ni lines will also allow scientists to determine whether this supernova was common envelope driven. Abundances can be a powerful engine diagnostic.

\begin{comment}
Discussion from CE meeting (put here for summary/help keep everyone in loop):

1) demonstrated neutron deficient nucleosynthesis in accreting neutron star common envelope - see throughout infall of neutron star into companion even when accreting material is helium-4 rich 
2) demonstrated that nucleosynthesis dependent on formation of accretion disk which leads to extended time for nucleosynthesis, see hot CNO cycle before peak of trajectory 

notes from discussion:
- Chris will work on intro to set up this story of angular momentum 
- in trajectory discussion should emphasise difference between Keegans and this work
- Christian will take charge of abstract and conclusions
- Alison and Christian would like paper submitted in next week so can add to grant app 
    
\end{comment}

\section{Acknowledgements}
ADHS, SEDA, AML and CAD acknowledge support from the Science and Technology Funding Council through grants ST/V001035/1 and ST/Y000285/1. The work by CLF and SWJ  was supported by the US Department of Energy through the Los Alamos National Laboratory. Los Alamos National Laboratory is operated by Triad National Security, LLC, for the National Nuclear Security Administration of U.S.\ Department of Energy (Contract No.\ 89233218CNA000001). The authors thank the NuGrid team for useful discussions and code support. LA-UR-26-21365. 

\bibliographystyle{mnras}
\bibliography{bibliography}

\appendix
\section{Additional Figures}

\begin{figure}
    \centering
    \includegraphics[width=\linewidth]{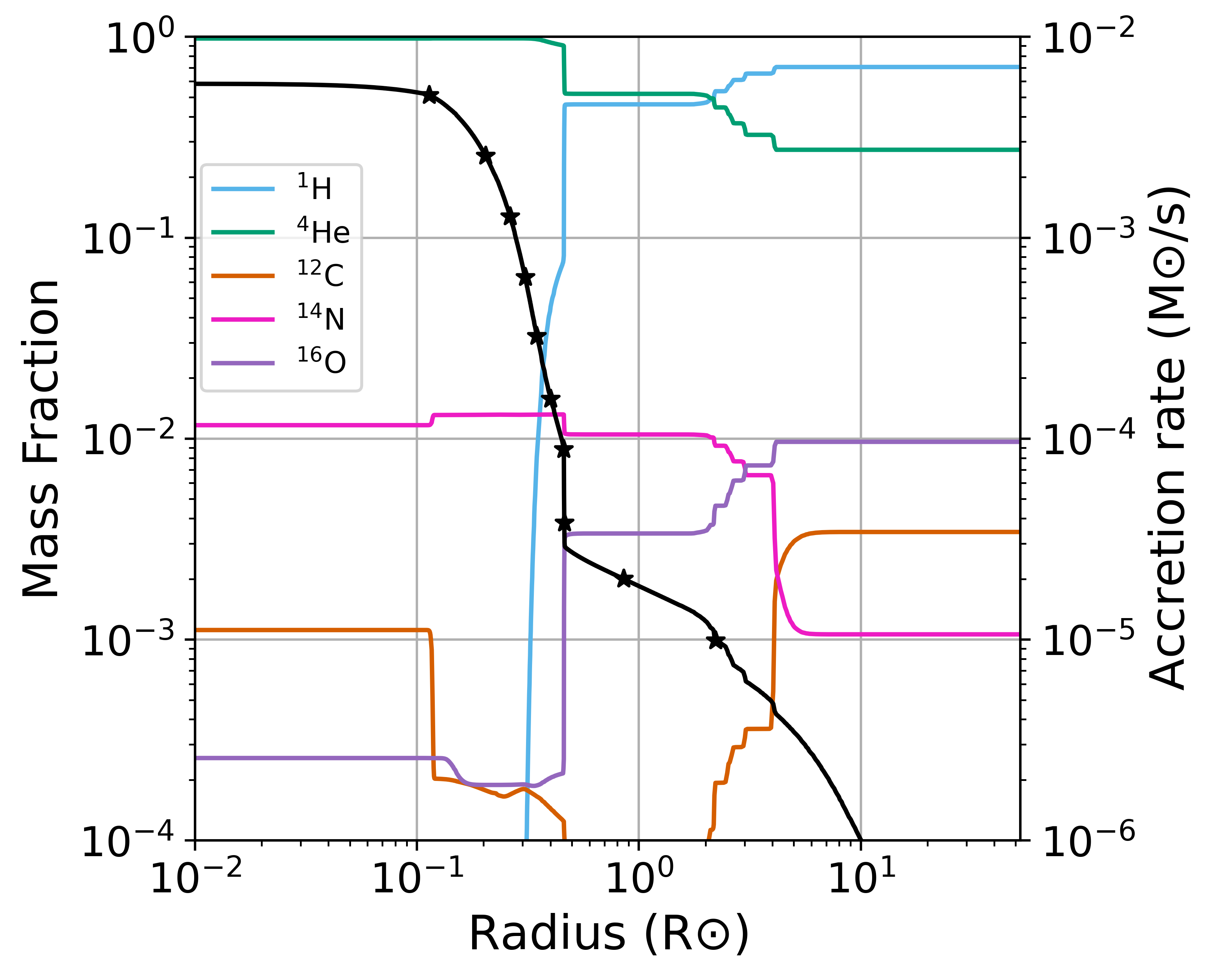}
    \caption{Composition of a 15 M$_{\odot}$ ZAMS star, when it has expanded to 52.4 R$_{\odot}$. The mass fraction (left axis) for $^{1}$H, $^{4}$He, $^{12}$C, $^{14}$N and $^{16}$O shown as a function of radial distance from the companion core. Also shown is the accretion rate (right axis) in solar masses per second vs radial distance. Black asterisks indicate accretion rates used in current study.}
    \label{fig:H_frac_to_iniab_mid}
\end{figure}

\begin{figure}
    \centering
    \includegraphics[width=\linewidth]{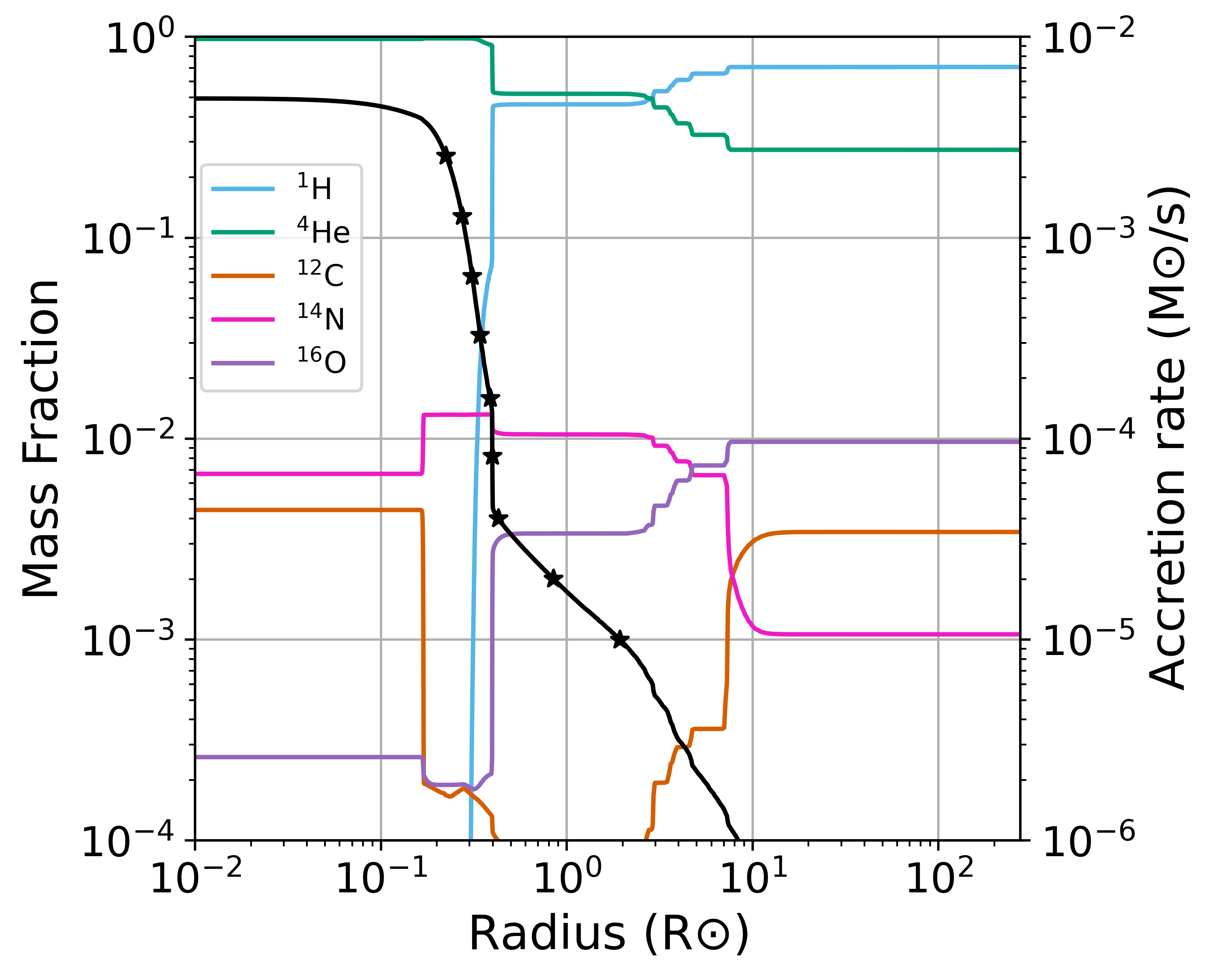}
    \caption{Composition of a 15 M$_{\odot}$ ZAMS star, when it has expanded to 275.6 R$_{\odot}$. The mass fraction (left axis) for $^{1}$H, $^{4}$He, $^{12}$C, $^{14}$N and $^{16}$O shown as a function of radial distance from the companion core. Also shown is the accretion rate (right axis) in solar masses per second vs radial distance. Black asterisks indicate accretion rates used in current study.}
    \label{fig:H_frac_to_iniab_late}
\end{figure}

Fig. \ref{fig:H_frac_to_iniab_mid} shows the mass fraction of $^{1}$H, $^{4}$He, $^{12}$C, $^{14}$N and $^{16}$O as a function of the 15 M$_{\odot}$ companion radius when the star has expanded to 52.5 R$_{\odot}$. The accretion rate as a function of companion radius is also shown. From this the initial composition for each accretion rate in each environment can be accurately calculated. Fig. \ref{fig:H_frac_to_iniab_late} shows the same as Fig. \ref{fig:H_frac_to_iniab_mid} but later in the companion evolution, when the companion radius has expanded to 275.6 R$_{\odot}$. 

%%%%%%%%%%%%%%%%%%%%%%%%%%%%%%%%%%%%%%%%%%%%%%%%%%

% Don't change these lines
\bsp	% typesetting comment
\label{lastpage}
\end{document}